\documentclass[conference]{IEEEtran}
\IEEEoverridecommandlockouts
\usepackage{cite}
\usepackage{amsmath,amssymb,amsfonts}
\usepackage{algorithmic}
\usepackage{graphicx}
\usepackage{textcomp}
\usepackage{xcolor}
\def\BibTeX{{\rm B\kern-.05em{\sc i\kern-.025em b}\kern-.08em
    T\kern-.1667em\lower.7ex\hbox{E}\kern-.125emX}}

\usepackage[caption=false,labelformat=simple]{subfig}
\usepackage{tabto}
\usepackage{longtable}
\usepackage{multirow}
\usepackage{booktabs}
\usepackage{array}
\newcolumntype{P}[1]{>{\centering\arraybackslash}p{#1}}
\usepackage{framed}
\usepackage{hyphenat}

\newenvironment{mybox}{\begin{oframed}\noindent}{\end{oframed}}

\newif\iflongver

\longverfalse

\newcommand{\todo}[1]{\textcolor{red}{TODO:#1}}

\begin{document}

\title{Adapting SQuaRE for Quality Assessment of Artificial Intelligence Systems
}

\author{\IEEEauthorblockN{Hiroshi Kuwajima}
\IEEEauthorblockA{\textit{DENSO CORPORATION} \& \textit{Tokyo Institute of Technology}\\
Tokyo, Japan \\
hiroshi.kuwajima.j7d@jp.denso.com}
\and
\IEEEauthorblockN{Fuyuki Ishikawa}
\IEEEauthorblockA{\textit{National Institute of Informatics}\\
Tokyo, Japan\\
f-ishikawa@nii.ac.jp}
}

\maketitle

\begin{abstract}
More and more software practitioners are tackling towards industrial applications of artificial intelligence (AI) systems, especially those based on machine learning (ML). However, many of existing principles and approaches to traditional 
% software 
systems do not work effectively for the system behavior obtained by training not by logical design. In addition, unique kinds of requirements are emerging such as fairness and explainability. To provide clear guidance to understand and tackle these difficulties, we present an analysis on what quality concepts we should evaluate for AI systems. We base our discussion on ISO/IEC 25000 series, known as SQuaRE, and identify how it should be adapted for the unique nature of ML and \textit{Ethics guidelines for trustworthy AI} from European Commission. We thus provide holistic insights for quality of AI systems by incorporating the ML nature and AI ethics to the traditional software quality concepts.
\end{abstract}


%\todo{final check: see comments}
% ML components vs. ML models}
% dataset vs. data set
% use abbreviations AI, ML, DNN(?)
% use italic for the terms from SQuaRE and Ethics guidelines??}
% SQuaRE (大文字小文字区別せずで置換)

\section{Introduction}
There has been increasing effort for industrial applications of artificial intelligence (AI) systems. This is particularly driven by technical advance in machine learning (ML) techniques including deep learning. Quality, dependability, or trust of such AI systems has been attracting wide attention both from the technical and social aspects.

Traditionally, the ML community has focused on accuracy over the whole data set (often just given). However, it is necessary to have more granular and specific evaluations in terms of requirements, to be reflected to data design, as well as consideration on a variety of other quality aspects.
% Machine learning models are widely used in many systems, and their quality requirements and evaluation is currently one-dimensional, {\it i.e.} accuracy. Conventional systems and software were evaluated against wide variety of requirements specifications such as quality characteristics defined in ISO/IEC 25000 series of standards, but machine learning models are usually evaluated only in terms of accuracy or error over whole test data set. Some accuracy requirements specification is encoded in the test data set, but it is not explicitly and exhaustively defined for machine learning models. It is necessary to define requirements and evaluation of wide variety of quality indicators for industrial use of machine learning models in systems.

We naturally consider adopting existing principles for traditional systems. The ISO/IEC 25000 series provides a framework or set of models for evaluation of software product quality, known as \textit{SQuaRE (System and Software Quality Requirements and Evaluation)}~\cite{ISO25000}. Although SQuaRE provides useful insights, it will not be directly applied to ML-based AI systems as it is. The essential difference is that ML components, such as deep neural networks (DNN), consist of enormous parameters and are constructed from training data. The resulting component is black-box and unexplainable, implementing a large fuzzy function such as image recognition and anomaly detection. 
% MLモデルで実装された機能がそのままシステム全体の機能になっている（なる場合もある）ので、システムとコンポーネントの区別が難しい
%quality of ML-based AI systems and that of ML components used in these systems are not distinct.
Such functions implemented by ML components often directly constitute the core functions of the whole system, quality of which is thus affected by the nature of ML. In one survey, more than 40\% of the survey participants answered existing approaches do not work for quality assurance of ML-based AI systems~\cite{MyCESSERIP19}.

On the other hand, new requirements are emerging given the advance of AI systems as shown in \textit{Ethics guidelines for trustworthy AI} published by European Commission (hereafter just ``Ethics guidelines''). Unique requirements have been discussed including human rights under AI autonomy, fairness, and explainability given the high impact of AI systems on human activities as well as the unique ML nature.

Industrial practitioners are required to holistically examine the whole picture of quality evaluation for emerging AI systems. Our objective in this paper is to support them by identifying the necessary updates on SQuaRE at its conceptual level (called quality characteristic and sub-characteristic) for AI systems. Our analysis is conducted from two viewpoints: what should be modified and what should be added. The first point is investigated by checking which existing concepts in SQuaRE are invalidated by the unique nature of ML. The second point is investigated by checking how the Ethics guidelines should be reflected to SQuaRE. We thus provide  holistic insights incorporating the ML nature and AI ethics to SQuaRE from traditional software engineering.

% Quality requirements and evaluation of conventional systems and software have been studied for years, and they include various perspectives besides accuracy. The historic outcome of systems and software quality is a good starting point. On the other hand, artificial intelligence gains worldwide publicity, and there are several trends to declare ethical standards for them. 
% such as Ethics guidelines for trustworthy AI (European Commission) 
% and OECD Principles on Artificial Intelligence (G20).
%And thus, we need to investigate if these quality standards cover ethical standards for AI.

%In this work, we study and propose machine learning models' 
%quality requirement and evaluation  by examining ISO/IEC 25000 series of standards (SQuaRE) 
% as a historic outcome of system and software quality,
%and Ethics guidelines for trustworthy AI published by European Commission.

%\todo{write with citations if the story of the abstract is ok}

The remainder of this paper is organized as follows. We first introduce the targets of our analysis, SQuaRE and the Ethics guideline, and also discuss related work in Section \ref{sec:background}. After presenting the methodology of our analysis in Section \ref{sec:analysis}, we present the results of our analysis from two approaches in Section \ref{sec:result}. We conclude the paper with remarks for future perspective in Section \ref{sec:conclusion}.

\section{Background}\label{sec:background}

\subsection{SQuaRE}

\iflongver
In software engineering, system and software quality models have been studied.
One of the earliest work is an international standard  ISO/IEC 9126 Software engineering --— Product quality, issued in 1991. ISO/IEC 9126 classified software quality into six characteristics, Functionality, Reliability, Usability, Efficiency, Maintainability, and Portability. ISO/IEC 9126 was replaced its succeeding international standard ISO/IEC 25010.
%Systems and software engineering --- Systems and software Quality Requirements and Evaluation (SQuaRE) --- System and software quality models.
\fi

The standard series of ISO/IEC 25000 or SQuaRE provides a framework or set of models for evaluation of software product quality. The core of SQuaRE is hierarchical (tree-structured) definition of \textit{quality models}, \textit{characteristics}, and \textit{sub-characteristics}, which defines the concepts or terminology about what we should evaluate in 
% software 
systems. \textit{Quality measures} define how to quantitatively evaluate each quality sub-characteristic in the form of a mathematical formula.
\iflongver
Below is an example branch of the hierarchical structure.
\indent\begin{tabular}{llll}
\multicolumn{4}{l}{Quality model: \textit{Product quality}}\\
& \multicolumn{3}{l}{Quality characteristic: \textit{Reliablity}}\\
&& \multicolumn{2}{l}{Quality sub-characteristic: \textit{Maturity}}\\
&&& : \textit{MTBF}\\
\end{tabular}
\else
An example branch of a quality model, a characteristic, 
a sub-characteristic, and a quality measure are \textit{Product quality}, \textit{Reliability}, \textit{Maturity}, and \textit{Mean time between failure (MTBF)}.
\fi
The \textit{MTBF} measure is defined as a rate of \textit{Operation time} and \textit{Number of system/software failures that actually occurred}. Here these two elements of the measure are called \textit{Quality measure elements (QMEs)}.

In this paper, we discuss two of the top-level quality models about evaluating systems, specifically, \textit{Product quality model} and \textit{Quality in use model}. These two models consider the development-time and run-time evaluation of systems, respectively.
We use abbreviations \textit{for product} and \textit{for use} when we want to clarify which quality model a (sub-)characteristic belongs to, i.e., \textit{Product quality} or \textit{Quality in use}, respectively.

% \todo{probably we should mention the position of data quality}
% TODO: Why not \textit{Data quality} and \textit{IT service quality}.

\iflongver
The whole picture of the two models are illustrated in Fig.~\ref{fig:product_model}
 and Fig.~\ref{fig:inuse_model}, respectively.
Boxes with thick lines and bold fonts, 
boxes with thin lines with bold fonts, 
and boxes with thin lines and normal fonts in these figures
represent quality models, quality characteristics, 
and quality sub-characteristics, respectively.
\begin{figure*}[htbp]
  \centering
  \includegraphics[scale=0.35]{SQuARE_product_model.eps}
  \caption{Product quality model and its quality (sub-)characteristics}
  \label{fig:product_model}
\end{figure*}
\begin{figure}[htbp]
  \centering
  \includegraphics[scale=0.35]{SQuARE_inuse_model.eps}
  \caption{Quality in-use model and its quality (sub-)characteristics}
  \label{fig:inuse_model}
\end{figure}
% There are too many QMs to be illustrated in a figure, 
% therefore QMs are omitted in Fig.~\ref{fig:product_model}
% and Fig.~\ref{fig:inuse_model}.
\else
% \todo{write a summarized note for quality models if necessary}
\fi

\subsection{Ethics Guidelines for Trustworthy AI}\label{ssec:ethics}
Ethics guidelines for trustworthy AI (\textit{Ethics guidelines}) were written by High-Level Expert Group on AI (AI HLEG), an independent expert group that was set up by the European Commission (EC) in June 2018.
The guidelines were published on April 8th, 2019, following the first draft released on December 18th, 2018.
The guidelines have four ethical principles: 
1) \textit{Respect for human}, 2) \textit{Prevention of harm}, 3) \textit{Fairness}, and 4) \textit{Explicability};
and seven key \textit{(ethical) requirements}.
% The relationship between ethical principles and requirements are not described in the guidelines. 
The guidelines also have a pilot assessment list that includes concrete assessment items associated to requirements.

\begin{figure*}[htbp]
  \centering
  \includegraphics[scale=0.35]{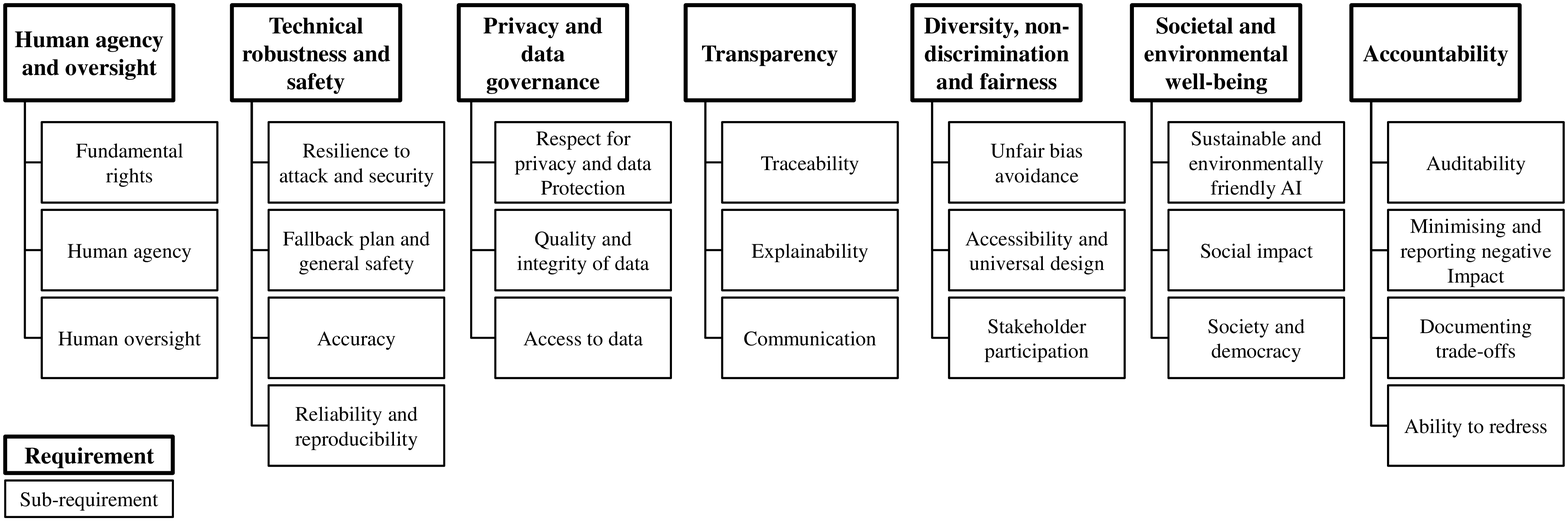}
  \caption{Seven Key Requirements in \textit{Ethics guidelines for trustworthy AI}}
  \label{fig:eu_ethics}
\end{figure*}

The list has a tree structure and the top-level requirements are 
1) \textit{Human agency and oversight}, 2) \textit{Technical robustness and safety}, 3) \textit{Privacy and data governance}, 4) \textit{Transparency}, 5) \textit{Diversity, non-discrimination and fairness}, 
6) \textit{Societal and environmental well-being}, and 
% \iflongver % DEBUG
7) \textit{Accountability}, as illustrated in Fig.~\ref{fig:eu_ethics}. 
% \else
% 7) \textit{Accountability}.
% \fi
The list also defines subcategories of the requirements.
We call them as \textit{sub-requirements}.
% \iflongver
Boxes with thick lines %and bold fonts, 
and those with thin lines %with normal fonts 
in Fig.~\ref{fig:eu_ethics} represent requirements and sub-requirements, respectively.

% \fi

Each sub-requirement includes a hierarchical check-list that consists of assessment items in the form of questions.
An example branch of a requirement, a sub-requirement, and an assessment item is \textit{Technical robustness and safety}, \textit{Resilience to attack and security}, and ``\textit{Did you assess potential forms of attacks to which the AI system could be vulnerable?}''
\iflongver
A child assessment item for the assessment item is ``\textit{Did you consider different types and natures of vulnerabilities, 
such as data pollution, physical infrastructure, cyber-attacks?}''
\fi
% There are too many assessment items to be illustrated in a figure, 
% therefore assessment items are omitted in Fig.~\ref{fig:eu_ethics}.

\subsection{Related Work}
Leading companies such as Google have published problem statements and guidelines based on their experience regarding testing (quality evaluation) or maintenance (technical debts)~\cite{Zinkevich-NIPS16wild-MLRules,Breck-NIPS16wild-MLTestScore,Sculley-SE4ML14-MLDebt}. The ML community summarized challenges in the data-centric nature, e.g., data management~\cite{Polyzotis-SIGMOD17-DataMgmtChallenges} and security~\cite{Liu-IEEEAccess18-MLSecurity}. There have been notably active discussions for explainability~\cite{Gunning-XAI}, adversarial examples~\cite{Goodfellow-ICLR15-AdversarialExamples}, and fairness~\cite{Davies-MLFairness}.

The research community of software engineering has been actively moving towards principles and techniques for ML-based AI systems. However, the initial outcome is almost limited to testing and verification techniques \cite{Zhang-MLTestingSurvey}.

The industrial practitioners need to work holistically on various aspects of quality. The objective of this paper is to provide clear guidance from this viewpoint, which will complement the above focused studies and support planning, decision making, and management activities in the industry.

\iflongver
\todo{ISO/JTC1/SC42?}
\todo{IEEE P7000?}
\todo{OECD guildelines?}
\todo{China guildelines?}
\fi

\section{Analysis Methodology}\label{sec:analysis}
We analyze the latest standards of SQuaRE series to identify how we should adapt them for ML-based AI systems, and how they cover Ethics guidelines for trustworthy AI. Specifically, we analyze what should be modified and what should be added, respectively, as depicted in the following research questions.
\begin{description}
\item[RQ1] How should we extend the existing (sub-)characteristics in SQuaRE when applied to ML-based AI systems?
\item[RQ2] What quality (sub-)characteristics should we add to SQuaRE for AI ethics?
\end{description}

For RQ1, we exhaustively check the existing QMEs, at the concrete level, to discover metrics that are not useful or that are not applicable when we consider ML-based AI systems. By focusing on the concrete metrics, we try to identify the gaps even though the quality characteristics or sub-characteristics are defined to be generic with abstract, broad terms.

For RQ2, we exhaustively check the assessment items in the Ethics guidelines. We attempt to map all the sub-requirements to the quality (sub-)characteristics, thus identifying what are essentially missing in SQuaRE.

\iflongver
In order to improve clarity in this paper, we visit our interpretation of terms in SQuaRE in Section~\ref{sect:interpretation}.
\fi

\section{Analysis Results}\label{sec:result}

\subsection{Modifying Existing (Sub-)Characteristics in SQuaRE}
\label{sect:square_result}
We first present the results of analysis on the unique nature of ML for RQ1. We identified two kinds of driving forces to adapt SQuaRE and discuss each of them in Sections \ref{sect:large_target} and \ref{sect:missing_tech}, respectively.

\subsubsection{Metrics with Large Fuzzy Evaluation Targets}
\label{sect:large_target}

One driving force for adapting SQuaRE is that large fuzzy functions are implemented with ML, such as image recognition for detecting different kinds of objects in a variety of situations.
This point is different from traditional software systems where functions are originally decomposed and given different logical (case-by-case) specifications and implementations.
Due to the difference, QMEs that count ``successful'' elements do not work in a straightforward way for ML-based AI systems (specifically, functions, tasks, and contexts). 
\iflongver
The 
\else
Some examples of the
\fi
QMEs that are affected by this point are illustrated in Table~\ref{tab:counting_qme}.
\iflongver
\else
\begin{table*}[ht]
\centering
\caption{SQuARE measures using number of successful tasks, functions, and contexts (Partial)}
\label{tab:counting_qme}
\begin{tabular}{@{} P{1.5cm}P{2cm}P{2.5cm}P{2.5cm}p{7.5cm} @{}} \toprule
\begin{tabular}{c}Quality\\Model\end{tabular} & \begin{tabular}{c}Quality\\Characteristic\end{tabular} & \begin{tabular}{c}Quality\\Sub-characteristic\end{tabular} & \begin{tabular}{c}Quality\\Measure\end{tabular} & \begin{tabular}{P{8.5cm}}Quality Measure Element (QME)\end{tabular} \\ \midrule
\multirow{4}{*}{\begin{tabular}{c}Product\\quality\end{tabular}} & \multirow{2}{*}{\begin{tabular}{c}Functional\\suitability\end{tabular}} & \multirow{2}{*}{\begin{tabular}{c}Functional\\completeness\end{tabular}} & \multirow{2}{*}{\begin{tabular}{c}Functional\\coverage\end{tabular}} & Number of {\bf functions} missing \\ & & & & Number of {\bf functions} specified \\ \cmidrule{3-5}
% & & \multirow{2}{*}{\begin{tabular}{c}Functional\\correctness\end{tabular}} & \multirow{2}{*}{\begin{tabular}{c}Functional\\correctness\end{tabular}} & Number of {\bf functions} that are incorrect \\ & & & & Number of {\bf functions} considered \\ \cmidrule{3-5}
% & & \multirow{4}{*}{\begin{tabular}{c}Functional\\appropriateness\end{tabular}} & \multirow{4}{*}{\begin{tabular}{c}Functional\\appropriateness of\\usage objective\end{tabular}} & Number of {\bf functions} missing or incorrect among those that are required for achieving a specific usage objective \\ & & & & Number of {\bf functions} required for achieving a specific usage objective \\
\cmidrule{2-5}
 & \multirow{2}{*}{\begin{tabular}{c}Maintainability\end{tabular}} & \multirow{2}{*}{\begin{tabular}{c}Testability\end{tabular}} & \multirow{2}{*}{\begin{tabular}{c}Test function\\completeness\end{tabular}} & Number of test {\bf functions} implemented as specified \\ & & & & Number of test {\bf functions} required \\ \cmidrule{1-5}
\multirow{4}{*}{\begin{tabular}{c}Quality\\in-use\end{tabular}} & \multirow{2}{*}{\begin{tabular}{c}Effectiveness\end{tabular}} & \multirow{2}{*}{\begin{tabular}{c}N/A\end{tabular}} & \multirow{2}{*}{\begin{tabular}{c}Tasks\\completed\end{tabular}} & Number of unique {\bf tasks} completed \\ & & & & Total number of unique {\bf tasks} attempted \\
\cmidrule{2-5}
 & \multirow{2}{*}{\begin{tabular}{c}Context\\coverage\end{tabular}} & \multirow{2}{*}{\begin{tabular}{c}Context\\completeness\end{tabular}} & \multirow{2}{*}{\begin{tabular}{c} Context\\completeness\end{tabular}} & Number of {\bf contexts} with acceptable usability and risk \\ & & & & Total number of required distinct {\bf contexts} of use \\
 %\cmidrule{3-5}
% & & \multirow{4}{*}{\begin{tabular}{c}Flexibility\end{tabular}} & \multirow{4}{*}{\begin{tabular}{c} Flexible context\\of use\end{tabular}} & Number of additional {\bf contexts} in which the product can be used with acceptable quality in use \\ & & & & Total number of additional {\bf contexts} in which the product might be used \\
\bottomrule 
\end{tabular}
\end{table*}
\fi

Now we particularly focus on an example sub-characteristic
\textit{Functional completeness}
% the degree to which the set of functions 
% {\it covers} all the specified tasks and user objectives.
that is the fraction of 
1) \textit{Number of functions missing}
and 2) \textit{Number of functions specified}, 
as illustrated in Table~\ref{tab:counting_qme}.
%In the context of machine learning, 
The following discussion can be generalized to 
other (sub-)characteristics using number of successful 
functions, tasks, and contexts.
Being evaluated for ML-based AI systems,
we interpret that 
functions ``missing'' are the functions 
that were not successfully trained,
even though developers specified to train them.
In other words, 
an ML-based AI system 
under evaluation fails on these missing functions,
even though developers 
specified the training or test data sets 
for the ML components used in the system
so that the data sets include such functions.
On the other hand, we interpret that functions ``specified'' are the functions 
that are included in such data sets, as well. 
Then, in order to measure 
\textit{Functional completeness}, 
we must count such functions in a data set.
However in an ML-based AI system,
a large fuzzy function is specified by a large data set
for the ML components used in the system,
{\it e.g.} a function to detect any objects 
in a data set with a certain 
dataset wide accuracy. 
Measuring \textit{Functional completeness} does not make sense for 
such a fuzzy large function.
%
% 以下は下の方に移動
% Second, the functions that were not successfully trained 
% reflect the training data set.
% In the context of machine learning systems,
% we interpret that the specified functions are included in 
% the training data set.
% and the criteria of achievement is not clear.

% Functional completenessだけに着目のため省略
% % Functional correctnessの説明
% On the other hand, Functional correctness is 
% the fraction of 1) total specified functions 
% and 2) total minus the incorrect functions.
% The same question rises here, {\it i.e.,} we do no not have 
% general criteria for correct and incorrect functions.
% Missing functions and incorrect functions 
% are not distinguishable in machine learning systems.
% % because machine learning models cannot achieve $100\%$ 
% % accuracy anyway.
% Therefore, for machine learning systems, 
% the boundaries between Functional completeness and Functional correctness 
% are blur, or not meaningful.
% Functional correctness become more important, 
% because it is also a characteristic of machine learning models 
% used in machine learning systems that they normally do not 
% achieve $100\%$ accuracy.

% Decomposeするという提案
% In the case of a quality measure Functional coverage, 
% Numbers of functions missing is not measurable,
% we cannot verify that each function using machine learning models 
% is missing (not achieved) or not.
In order to measure such QMEs 
for a large fuzzy function trained in a data driven manner,
we decompose it into 
fine-grained functions~\cite{Czarnecki-WAISE18-PerceptualUncertainty} 
% We %do not assume natural distributions of data, but do 
by splitting a data set into 
partitions that include small functions.
The decomposition must be done in an application-specific way,
greatly incorporating domain experts and their domain knowledge.
ML-based AI systems will be evaluated for each partition, 
and the functions
associated with that partition 
are counted for quality measurement.
% These are in the following two steps: 
% \begin{enumerate}
%     \item Requirements specifications define fine-grained tasks, functions, and contexts that test data set must include.
%     \item Requirements specifications are verified by performing better accuracy in the test data set than specified.
% \end{enumerate}
% 以下は必要？
In the case of pedestrian detection, we can consider more meaningful evaluation by splitting the data set to capture specific types of pedestrian and weather condition (functions and contexts).
%wide variety of pedestrians and weathers, etc. should be specified in the requirements or design specification to split data sets into partitions that include small functions, and contexts, that is, a data set is split into each partition that has specific types of pedestrian and weather observed.
% Appropriateness of granularity can be an extended quality.
% Example of the criteria for granularity are ...
% Data partitioning is applicable to other quality measures 
% illustrated in Table~\ref{tab:counting_qme}.

\begin{mybox}
\textbf{Extension A1: Decomposition of Evaluation Target}\\
(Sub-)characteristics to measure number of successful tasks/functions/contexts, {\it i.e}, a characteristic \textit{Effectiveness} (\textit{Quality in use}) and sub-characteristics \textit{Functional completeness}, \textit{Functional correctness}, \textit{Functional Appropriateness}, \textit{Testability} (\textit{Product quality}), \textit{Context completeness}, \textit{Flexibility} (\textit{Quality in use}), should be extended to consider application-specific decomposition of a large fuzzy task/function/context implemented by ML components.
\end{mybox}

% Functional appropriatenessの説明
% これはあまり機械学習には関係ないので省略しても良い。
% Although aforementioned Functional completeness and 
% Functional correctness are 
% measured based on specified functions,
% Functional appropriateness is task oriented.
% That is, Functional appropriateness 
% is measured based on tasks or their objectives instead of functions.
% Important functions that are repeatedly required for 
% achieving multiple tasks and objectives are highly weighted 
% for Functional appropriateness.
% Functional appropriateness is a mixture of Functional completeness 
% and Functional correctness.
% It does not discriminate missing functions and incorrect functions.

% どのデータ・セットで数えるのか？
Second, if a large fuzzy function is decomposed by splitting data sets, 
then selection of data sets to split is important.
% 機械学習はテストデータで検証するだけで仕様が無い
In the current practice, 
 an ML components is evaluated 
 as the total performance on a test data set.
However, \textit{Functional completeness} is not the accuracy
for a given test data set but for a given specification,
and it should be extended 
to connect data sets and specifications.
Collecting training, test, and operational data 
are (part of) designing, 
building requirements,
and investigating actual usage of ML components, respectively.
Thus, 
the functions
specified by design specifications and those specified 
by requirements specifications should be included in 
the training and the test data, respectively.
Those counted for use should be included
in the operational data.
For example in a measure \textit{Task completed}, 
unique tasks completed and those attempted are the successful tasks 
and the tasks attempted both in operational data, respectively. 
%In \textit{Context completeness} measure,
%contexts with acceptable usability and risk, 
%and required distinct contexts of use are 
%interpreted as successful contexts in operational data,
%and contexts included in the test data sets, respectively.
%In \textit{Flexible context of use} measure,
%additional contexts with acceptable quality in use and possible additional contexts
%are successful contexts in operational data but not in test data sets and
%are those including unsuccessful ones.

\begin{mybox}
\textbf{Extension A2: Quality Measure and Data Set}\\
% If application-specific decomposition of a large task/function/context is done by decomposing data sets in the characteristics and sub-characteristics to evaluate tasks/functions/contexts, then it should be taken into consideration that training, test, and operational data sets have different roles. 
(Sub-)characteristics to measure successful tasks/functions/contexts should be extended to consider the relationships to training, test, and operational data sets that have different roles for ML components.
\end{mybox}

% 数えるときに、正解不正解は0/1ではわからない。
% 不確実性を入れるべきでは？という主張。
Third, we need to count ``successfully trained'' functions.
It is not simple 
for ML-based AI systems, because ML components 
normally do not achieve $100\%$ accuracy.
ML components can success or fail on the same function, {\it i.e.}, 
we must handle the uncertainty of measurement,
even if we get fine-grained functions.
The behavior of ML components, DNN in particular, changes unstably, because they are highly nonlinear.
It should be considered that uncertainty indices such as bias and variance are added for each quality measure.

\begin{mybox}
\textbf{Extension A3: Handling of Uncertainty}\\
(Sub-)characteristics to measure number of successful tasks/functions/contexts should be extended to consider the uncertainty of measurement.
\end{mybox}

% 長過ぎる、上と同じようなことを言っている。
% しかし、境界値が引けないこと、内挿性が無いこと、透明性がないこと、は手短に考察として書きたい。
% \todo{too long?} 
% Discussion for Ex-A1, need section?
% \subsubsection*{Discussion for Ex-A1}
Extension A1-3 address the following inabilities of an ML component: 
% 1) unable to interpret the internals,
1) It does not have specification and functions are not explicitly specified;
2) It is highly nonlinear and the behavior is not robust even within a small function.
These properties are important for quality measurement for conventional software. 
% First, we can expect interpretability in conventional software, 
% because it is manually designed and its complexity is handled by humans. 
% Such interpretabile property of conventional software 
% greatly facilitated the ease of quality evaluation, {\it e.g.} white box testing.
% In machine learning models, or deep learning models in particular, 
% numerous number of parameters are automatically trained by optimizer 
% in a data-driven manner, and humans cannot interpret the resulted models. 
% We trade performance in return for lack of interpretability; 
First, we implement conventional software 
based on specifications, and thus, 
we are able to identify functions based on them. 
% However we cannot define partitions for machine learning models, 
% because we have training and test data sets but no specification, 
% and machine learning models are automatically trained based on training data sets.
Then, conventional software can be evaluated for each function.
% assumed discrete tasks/objectives/functions based on specifications,
% and quality evaluation was conducted on each of them. 
% However, unlike conventional software, 
% machine learning models cannot have specifications; 
Second, robustness is prerequisite for quality evaluation based on specifications. 
In conventional software, the behavior of a function is robust, therefore it can be treated as a single unit of quality evaluation.
% Toward the quality evaluation of machine learning models in the same manner of conventional software,
% we need to tackle challenges of specifications and testability of machine learning models.
% A1 is toward the specifications of machine learning models.
% A3 is toward the testability of machine learning models.

\iflongver
% SQuAREの拡張提案ではないので。
One future work is to generate specifications posteriori for a trained ML components; 
it shall be an agile (iterative) process between training and specification. 
It is true that we cannot provide requirements or design specifications for an ML component in advance to training it.
A part of future work is to train ML components with good properties for quality evaluation such as smooth behavior (high robustness) and clear partitions, as well as high performance; another part is to generate specifications after training. 
These challenges are toward testability and specifications of ML components, respectively.
% end of Ex-A1
\fi

% 以下は倫理要求から来るということにして、ここからは削除。
% \input{corner_case_measure.tex}
% Table~\ref{tab:corner_case_qme} illustrates 
% the quality measures that can involve corner cases.
% These quality measures are the type of SQ1;
% they can be extended by effort based ones,
% such as time consumed to explore corner cases 
% during development time.

\subsubsection{Metrics with Missing Techniques}
\label{sect:missing_tech}

The other kind of driving force to adapt SQuaRE is that there are sub-characteristics for which effective measurement technique has not been established in the case of ML-based implementation, especially DNN.

A quality sub-characteristic \textit{Operability}
has a measure 
\iflongver
\textit{Monitoring capability}, 
as illustrated in Table ~\ref{tab:change_qme}.
\else
\textit{Monitoring capability}.
\fi
Explainable AI (XAI)~\cite{Gunning-XAI} 
is a rapid growing area in the artificial intelligence research, 
and techniques to explain or interpret ML components 
are proposed in recent
% \iflongver
years~\cite{Ribeiro-KDD16-ExplainClassifier,Kuwajima2019}.
% years~\cite{Kuwajima2019}.
% \else
% years.
% \fi
They can be used for monitoring capacities for 
ML-based AI systems, but XAI research is still in 
the very early stage. 
Recommended techniques of XAI have yet to be developed.

\begin{mybox}
\textbf{Extension A4: Investigation of Monitoring Capacity}\\
A sub-characteristic \textit{Operability} (\textit{Product quality}) that has a measure \textit{Monitoring capacity} should be extended with recommended monitoring (XAI) techniques.
\end{mybox}

It has been reported that image
recognition models incorrectly infer classes or objects with high confidence, due to small noise that cannot be recognized by humans.
Such noise is called adversarial examples (AEs)~\cite{Goodfellow-ICLR15-AdversarialExamples},
and can be a new type of data corruption 
for ML components.
Corruption prevention methods against AEs are studied in recent years,
but we do not have the established or recommended techniques until now.
\iflongver
In Table ~\ref{tab:change_qme}, a
\else
A
\fi
sub-characteristic \textit{Integrity} that has a measure 
\textit{Internal data corruption prevention} should 
address such AEs in addition to the conventional data corruption.

\begin{mybox}
\textbf{Extension A5: Investigation of Data Integrity}\\
A sub-characteristic \textit{Integrity} (\textit{Product quality}) 
should be extended to evaluate the impact of 
ML specific data corruption such as 
adversarial examples.
% Metrics should be extended with available recommended techniques.
\end{mybox}

SQuARE defines a characteristic \textit{Maintainability}
that has a measure 
\iflongver
\textit{Coupling of components}
as illustrated in Table ~\ref{tab:change_qme}.
\else
\textit{Coupling of components}.
\fi
However, coupling of ML components 
is not well studied, that is, we do not know whether a component has no impact on others. If an ML component is trained along with specific surrounding components and systems, then, it is tightly fit them and cannot be decoupled. However, we cannot measure such modularity with current technologies.

\begin{mybox}
\textbf{Extension A6: Investigation of Modularity}\\
A sub-characteristic \textit{Modularity} (\textit{Product quality}) should be extended to evaluate the expected ML components' independency from surrounding components and systems.
\end{mybox}

Work time to make a specific type of modification is 
also an important aspect of maintainability,
and that is defined as 
a measure \textit{Modification efficiency} in a characteristic \textit{Maintainability}.
That work time for ML-based AI systems should include the training time to make a specific type of 
modification for ML components.
However, expected work time, {\it i.e.}, expected training time is 
generally unknown in advance to training with current technology.
% The maintainability of the model is low in these cases, 
% but we cannot measure its modularity and modifiability.
%and the measure needs to be changed, or cannot be used.
% These quality measures are the type of SQ1.

\begin{mybox}
\textbf{Extension A7: Investigation of Modifiability}\\
A sub-characteristic \textit{Modifiability} (\textit{Product quality}) should be extended to evaluate the expected training time of ML components used in ML-based AI systems, without actually conducting training. 
\end{mybox}

The complexity of systems is one of key factors of maintainability.
In SQuARE, cyclomatic complexity score is used in a measure
\textit{Cyclomatic complexity adequecy}. It is represented by the number of linearly independent execution paths,
and is not suitable for ML components, 
because they have only one linearly independent path and the score always equals to a constant.
ML components can show different complexity in other manners.
Number of parameters and layers, numerical precision 
like 8-bit unsigned integer and 16-bit floating point, 
or simply number of multiply\hyp{}accumulate operations (MACs)
performed
can be used to measure the structural (or computational) 
complexity of ML components.
In addition to such structural complexity,
we need the behavioral complexity, {\it i.e.} robustness, 
for ML components.
ML components with the same 
structure (architecture) but with different parameters 
can behave differently;
a component can show highly nonlinear behavior 
but another can show robust behavior.

\begin{mybox}
\textbf{Extension A8: Investigation of Complexity}\\
A sub-characteristic \textit{Maintainability} (\textit{Product quality}) should be extended to evaluate both structural complexity and behavioral complexity (robustness) of ML components.
%that have only one linearly independent execution path. 
\end{mybox}

% The quality measures to cover the Ethics guidelines for trustworthy AI 
% become important, that is, those are the type of SQ3;
% we will discuss such quality measures in Section~\ref{sect:guideline_result}.

\subsection{Adding New (Sub-)Characteristics in SQuaRE}
\label{sect:guideline_result}
We present the results of the second analysis on the Ethics guidelines for RQ2. We exhaustively checked all the sub-requirements in the Ethics guidelines, map all them items to the quality (sub-)characteristics in SQuaRE, and identified extensions to SQuaRE for AI systems.

\subsubsection{Autonomy and Human}
The sub-requirement of \textit{Fundamental rights} includes assessment items on negative impacts on fundamental rights, unintended interference on human decision making, and notification about existence of non-human agents. This sub-requirement basically reflects the respect on human autonomy.

There are four sub-requirements about interaction between human and AI systems in the Ethics guideline. Specifically, \textit{Human agency} and \textit{Human oversight} mention prevention of overconfidence and appropriate control by human, respectively. The sub-requirement of \textit{Explainability} is about user understanding the decision and outcome by AI systems. The sub-requirement of \textit{Communication} also mentions similar points but put more focus on clarification for the target audience, feedback cycles, and psychological aspects.

Among the characteristics for use in SQuaRE, the quality characteristic of \textit{Freedom from risk} matches with this aspect. Currently, sub-characteristics regarding economic risk, health and safety risk, and environmental risk are included. It is natural to add a sub-characteristic about risk on human rights.

\begin{mybox}
\textbf{Extension B1: Risk on Human Autonomy}\\
A sub-characteristic \textit{Mitigation of risk on human autonomy} should be added in the \textit{Freedom from risk} characteristic (\textit{Quality in use}).
\end{mybox}

Regarding the characteristics for product, we interpret all of the above aspects are extending the traditional notion of usability; now human not only use systems by their commands but allow and rather expect autonomous systems to run proactively but still under control and understanding. We therefore propose to extend the characteristic of \textit{Usability} to incorporate this change.

\begin{mybox}
\textbf{Extension B2: Collaboratability}\\
A characteristic \textit{Usability} should be extended to \textit{Collaboratability} to reflect autonomous roles of AI systems. Additional sub-characteristics should include \textit{Controllability} and \textit{Explainability} as well as \textit{Collaboration Effectiveness} (\textit{Product quality}).
\end{mybox}

\subsubsection{Fairness}
The sub-requirement of \textit{Unfair bias avoidance} mentions the demand for avoiding unfair bias or considering diversity of users. As this point has been one of the key issues for ML-based AI systems~\cite{Davies-MLFairness}, we extend SQuaRE to include it. In the Ethics guideline, one description of fairness refers to equal and just distribution of both benefits and costs as well as freedom from unfair bias, discrimination, and stigmatisation. This point is included in parallel with the \textit{Extension B1} as a different kind of risks.

% \todo{also in product quality??}

\begin{mybox}
\textbf{Extension B3: Fairness}\\
A sub-characteristic \textit{Mitigation of risk by unfair bias} should be added in the \textit{Freedom from risk} characteristic (\textit{Quality in use}).
\end{mybox}

\subsubsection{Accuracy}
The requirement \textit{Technical robustness and safety} describes technical points that have been basically common for dependable systems. Differences in AI systems are the focus on \textit{Accuracy} (one of the sub-requirement).
% TODO: as well as different kinds of attacks mentioned in the assessment items such as data pollution and adversarial attacks.
Capturing this point has been discussed as one of the core topics in Section \ref{sect:square_result} regarding evaluations of functionality, e.g., completeness and correctness.
% TODO: too rough
% \todo{focus on accuracy: covered by Ax? Make consistent and more explicit.}
% \todo{adv. example: by A3?}
% \todo{data pollution: out of the scope as internal aspects?}

\subsubsection{Privacy}
The sub-requirement \textit{Respect for privacy and data protection} mentions data protection, minimal use of sensitive or personal data, control over personal data, and other similar issues. Surprisingly, SQuaRE did not include specific items for privacy though some parts are covered by the \textit{Security} characteristic. This is probably because the demand for privacy recently emerged given increasing use of data.

We should naturally extend SQuaRE to include privacy concerns, probably not limited to targeting AI systems.

\begin{mybox}
\textbf{Extension B4: Privacy}\\
The \textit{Security} characteristic should be extended to \textit{Security and Privacy}, to incorporate an additional sub-characteristic of \textit{Privacy} (\textit{Product quality}).
\end{mybox}

\subsubsection{Accountability}
The sub-requirement of \textit{Reliability and reproducibility} includes, among the common concepts of reliability, the concept of reproducibility, that is, whether the same behavior can be exhibited in experiments with the same condition. This is a point attracting wide attention in AI research where outcomes may be affected by randomness and configuration parameters.

The sub-requirement of \textit{Traceability} mentions documentation of how the system is constructed, e.g., algorithms and testing methods. This sub-requirement suggests the demand for responsibility in algorithmic decisions, as already included in GDPR (General Data Protection Regulation). The concept is different from the traditional notion of ``traceability'' in software engineering, which was about management of (often internal) deliverables for the purpose of maintainability. Indeed, traceablity appears as a note for the \textit{Reusability} sub-characteristic in SQuaRE.

The sub-requirement \textit{Auditability}, in the requirement \textit{Accountability}, mentions traceability and logging of processes and outcomes as well as separated auditability for each aspect.

Given the increasing demand for accountability, as in GDPR, we note the significance of these aspects. However, we interpret these aspects come at the meta-level of SQuaRE (sub-)characteristics, rather than a (sub-)characteristic.

\begin{mybox}
\textbf{Meta-Level Consideration: Accountability}\\
There is increasing demand for accounting the evidences that justify the evaluation results for (sub-)characteristics. This point should be noted when evaluation activities are planned and conducted.
\end{mybox}

\subsubsection{Other Requirements}
The sub-requirement of \textit{Fallback plan and general safety} mentions fallback plans and safety risks. The sub-requirement of \textit{Reliability and reproducibility} also mentions general aspects of reliability assessment, except for the reproducibility part. These aspects have been covered in the \textit{Reliability} and \textit{Security} characteristics for product as well as the \textit{Freedom from risk} characteristic for use.

The sub-requirements of \textit{Quality and integrity of data} and \textit{Access to data} mention data management, monitoring, access control, and so on. These aspects have been partially covered in the \textit{Security} characteristic but are most assessment items are about the internal implementations, which is out of the scope for this paper.
Data quality of SQuARE may cover them.

The sub-requirement of \textit{Accessibility and universal design} mentions consideration of disabilities and people from different backgrounds. This is included in the \textit{Accessibility} sub-characteristic for product.

The sub-requirement of \textit{Stakeholder participation} was considered as out of the scope. This is rather a recommendation on the process, not evaluation of the system itself, and such an aspect has not been included in SQuaRE. We also exclude the sub-requirements of \textit{Social impact} and \textit{Society and democracy} about social impacts of the AI systems.

In the analysis, we have excluded assessment items regarding organizational activities, such as insurance policy. For the same reason, the sub-requirements of \textit{Minimising and reporting negative impact}, \textit{Documenting trade-offs}, and \textit{Ability to redress} are out of the scope.

\section{Concluding Remarks}\label{sec:conclusion}
% SQuaREは次の改定で機械学習への対応をしたほうがいい。今回の検討では、その一例を示した。
% AI Ethics Guidelinesには具体的な評価方法が必要。今回の検討では、その一例を示した。
% 外部品質だけではなく、訓練データなどの内部品質も検討していくべきである。

In this paper, we have presented our analysis on how to adapt SQuaRE for ML-based AI systems. Obviously, the current version of SQuaRE did not take ML-based implementations into consideration as at that time ML was almost in laboratory. Nevertheless, not limited to the specific nature of ML, SQuaRE could take updates to reflect the increasing impacts of systems on human activities, for example to consider risks for human rights and privacy. We also reviewed the coverage of Ethics guidelines for trustworthy AI with SQuaRE, which revealed most part of the Ethics guidelines are not covered by SQuaRE. We believe this preliminary work provides proper guidance for industrial practitioners without waiting for the long-lasting update process of the standards. 
%We conducted change analysis on a system and software quality models SQuaRE, in order to apply them to machine learning systems. SQuaRE has been widely used for systems and software, and on the other hand, machine learning systems occur in the context of dramatic improvement of machine learning, in particular deep learning models. It is good for SQuaRE to follow this trend by incorporating the particular characteristics of machine learning systems. Our work is a preliminary attempt to it.

%he ethics guidelines have a (pilot) assessment list, it needs more specific measures in order to evaluate assessment items. 

Our focus in this paper has been so-called external quality of systems, not about internal quality about intermediate activities and deliverables through the process of development and operation. We will continue our investigation to consider internal quality aspects so that guidance is provided for concrete activities on training data, specification documents, test design, runtime monitoring, and maintenance.

%At last, we note that our analysis is only on the external quality, {\it i.e.}, product quality model and quality in-use model. However in developing good machine learning models used in machine learning systems, development data, that is training and test data sets, is important; they are corresponding to source codes of conventional software, and thus their quality is internal quality. Training data quality directly affects quality of machine learning models, that is, the connection between an internal quality and an external quality is more obvious than that of conventional software. Thus, a new internal quality model is desired for machine learning systems.

\bibliographystyle{IEEEtran}
\bibliography{references}

\end{document}